\documentclass[fleqn,usenatbib]{mnras}

\usepackage{newtxtext,newtxmath}
\usepackage{xspace}

\usepackage[T1]{fontenc}
\usepackage{booktabs} 
\usepackage{stfloats}

\DeclareRobustCommand{\VAN}[3]{#2}
\let\VANthebibliography\thebibliography
\def\thebibliography{\DeclareRobustCommand{\VAN}[3]{##3}\VANthebibliography}

\usepackage{graphicx}	
\usepackage{amsmath}	
\usepackage{orcidlink}
\usepackage{bm}
\usepackage{float}
\usepackage{savesym}
\savesymbol{tablenum}
\usepackage{siunitx}
\usepackage{ulem}

\newcommand{\kms}{$\mathrm{km\,s^{-1}}$}
\newcommand{\msun}{M_{\sun}}

\newcommand{\hi}{{H\:\sc{i}}\xspace}

\defcitealias{Paul2023}{P23}

\title[HI Intensity Mapping Model]{Modeling the Nonlinear Power Spectrum in Low-redshift \hi Intensity Mapping}

\author[Li et al.]{
Zhixing Li,$^{1,2}$\thanks{E-mail: zhixing.li-2@postgrad.manchester.ac.uk}
Laura Wolz,$^{2}$\thanks{E-mail: laura.wolz@manchester.ac.uk}
Hong Guo,$^{3}$\thanks{E-mail: guohong@shao.ac.cn}
Steven Cunnington$^{2}$
and Yi Mao$^{1}$
\\
$^{1}$Department of Astronomy, Tsinghua University, Beijing 100084, China\\
$^{2}$Jodrell Bank Centre for Astrophysics, Department of Physics and Astronomy, The University of Manchester, Manchester M13 9PL, UK\\
$^{3}$Shanghai Astronomical Observatory, Chinese Academy of Sciences, Shanghai 200030, China
}

\begin{document}
\label{firstpage}
\pagerange{\pageref{firstpage}--\pageref{lastpage}}
\maketitle

\begin{abstract}
We present a simulation-based framework to forecast the \hi power spectrum on non-linear scales ($k\gtrsim 1\ {\rm Mpc^{-1}}$), as measured by interferometer arrays like MeerKAT in the low-redshift ($z\leq 1.0$) universe. Building on a galaxy-based \hi mock catalog, we meticulously consider various factors, including the emission line profiles of \hi discs and some observational settings, and explore their impacts on the \hi power spectrum. While it is relatively insensitive to the profile shape of \hi emission line at these scales, we identify a strong correlation with the profile width, that is, the Full Width at Half Maxima (FWHM, also known as $W_{\rm 50}$ in observations) in this work. By modeling the width function of $W_{50}$ as a function of $v_{\rm max}$, we assign each \hi source a emission line profile and find that the resulting \hi power spectrum is comparatively close to results from particles in the IllustrisTNG hydrodynamical simulation. After implementing $k$-space cuts matching the MeerKAT data, our prediction replicates the trend of the measurements obtained by MeerKAT at $z\approx 0.44$, though with a significantly lower amplitude.  Utilizing a Monte Carlo Markov Chain sampling method, we constrain the parameter $A_{W_{\rm 50}}$ in the $W_{\rm 50}$ models and $\Omega_{\rm HI}$ with the MeerKAT measurements and find that a strong degeneracy exists between these two parameters.

\end{abstract}


\section{Introduction}
\label{sec:intro}

In a low-redshift universe ($z\leq 1.0$), more than 95 percent of atomic hydrogen (\hi) gas resides in the cold dense region of galaxies within dark matter halos, where it can be self-shielded from ultra-violet photons \citep[see e.g.,][]{Krumholz2009, Diemer2018, Villaescusa-Navarro2018}. The detectable 21~cm radiation emitted by the hyperfine transition of \hi makes it a novel and competitive probe to measure the matter distributions and structures in the post-Epoch of Reionization (post-EoR) Universe at various spatial scales.

In the very local Universe ($z\leq 0.1$), large dish radio telescopes such as Arecibo, Parkes and Five-hundred-meter Aperture Spherical radio Telescope (FAST) have detected 21 cm signals emitted by extragalactic sources \citep{Haynes2018, Barnes2001, Zhang2024}. The extragalactic \hi surveys performed by these telescopes have provided a wealth of combined information on \hi mass, redshift, velocity profile and yielded valuable insights into various aspects of \hi statistics, including the \hi\ Mass Function (HIMF) and the consequent \hi abundance $\Omega_{\rm HI}$ \citep[e.g.,][]{Martin2010, Jones2018}, the \hi-halo mass relation \citep{Guo2017, Guo2020}, etc. However, at higher redshifts ($ z \geq 0.1$), as the distances between the observer on Earth and the \hi sources continually increase, the emission lines become significantly fainter and harder to distinguish from the noise. To overcome this limitation, the stacking method and intensity mapping (``IM'' hereafter) technique is widely applied, both of which can measure the combined signal emitted by several unresolved \hi sources instead of resolving individual objects, but only IM retains the spatial information \citep[e.g.][]{Battye2004,Mao2008,Chang2008,Wyithe2009,Battye2013}. 

Numerous radio telescopes utilizing IM are currently either under construction or already in operation, such as the BAO from Integrated Neutral Gas Observations \citep[BINGO;][]{BINGO2019}, the Canadian Hydrogen Intensity Mapping Experiment \citep[CHIME;][]{CHIME2014}, the Square Kilometre Array \citep[SKA;][]{SKA2009} and its pathfinder MeerKAT \citep{Jonas2016}. These telescopes fall into two main categories: single-dish telescope and interferometer array. In the case of the MeerKAT and SKA-mid arrays, they can be used in both observation modes, providing flexibility. The single-dish mode is optimised to observe comparatively large patches and collect the overall signal from relatively fast scans. On the contrary, the interferometer mode captures the signals from smaller fields of view by combining multiple small dishes. With the baselines, i.e., the distances between dishes ranging from meters to kilometres, the interferometer mode can achieve higher resolution, allowing for detailed studies of non-linear scales.

Notably, \cite{Paul2023} presented the first measurements of the \hi power spectrum on non-linear scales obtained by the interferometer mode of MeerKAT, making it urgent to obtain a theoretical reference on these scales. Therefore, this work mainly focuses on the \hi power spectrum at non-linear scales, to provide a theoretical framework for upcoming and existing \hi IM surveys with inteferometers and improve the interpretation of the data.

When dealing with non-linear scales, the power spectrum of compact astrophysical objects such as galaxies or \hi sources is not solely influenced by the linear part. The shot noise term, which arises from the discrete characteristics of these compact objects, commonly behaves like a plateau and dominates as $k$ increases to non-linear scales \citep[e.g., see][]{Wolz2019}. Note that large scales in real space correspond to small $k$ modes in Fourier space, and vice versa. However, in the case of \hi, the significant internal movement of the \hi gas within the discs, including both circular rotation and random motion, results in the elongation of the \hi sources along the line of sight (LOS) due to the Doppler effect in redshift space. Therefore, since the \hi discs are no longer compact (though only in the LOS), the \hi power spectrum appears to continuously decrease with increasing $k$ compared to a constant shot noise term \citep{Sarkar2019, Zhang2020}.

Addressing the kinematics of the \hi gas is crucial when measuring the power spectrum on small scales. Previous studies using the particle data in hydrodynamical simulations show no `shot noise plateau' for \hi power spectrum in redshift space \citep{Villaescusa-Navarro2014, Villaescusa-Navarro2018}. Other studies have constructed some models for the \hi velocity profiles, confirming the suppression of power spectrum at small scales by the inner motion of the \hi discs \citep[see e.g.,][]{Sarkar2019,Zhang2020}. However, these models have yet to fully explore certain aspects, such as the choice of halo- or galaxy-based \hi distributions, reasonable \hi disc sizes, the impact of velocity profile shape on the \hi power spectrum and so on. The observational settings in the power spectrum measurements are also widely overlooked.

\begin{figure*}
	\centering
	\includegraphics[width=0.7\textwidth]{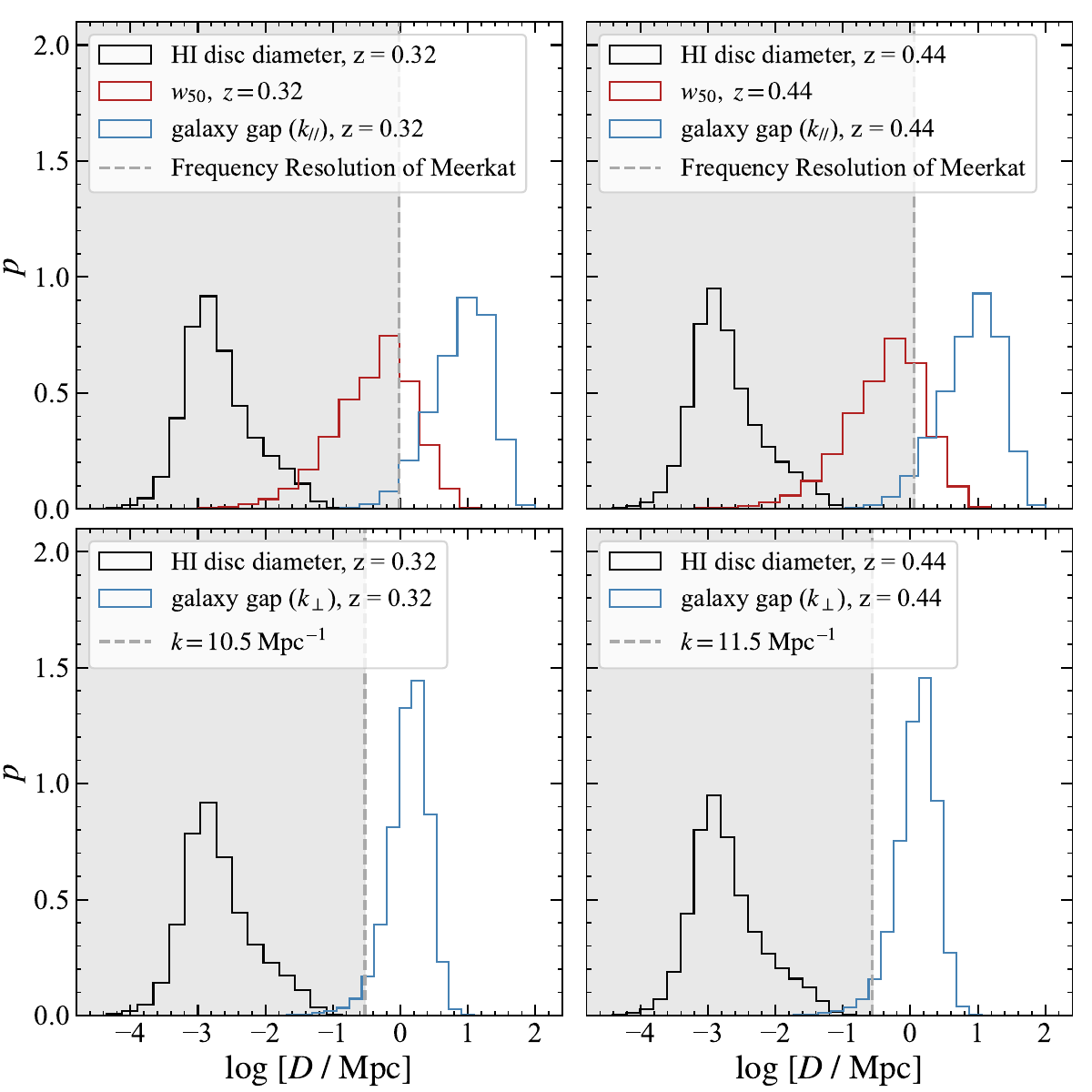}
	\caption{The probability distribution of the scales for \hi profile (black lines), \hi emission line width in different direction (red lines) and the gap between galaxies (blue lines) in $k_{||}$ direction (top panels) and in $k_{\perp}$ direction (bottom panels). The gray dashed lines in the top panels represent the frequency resolution, i.e., channel width adopted by the auto-correlation measurement (209kHz in $z=0.32, 0.44$). In the bottom panels, the gray dashed lines represent the scale corresponding to the largest $k$ modes ($k \lesssim 10.5\ {\rm Mpc}^{-1}$ at $z=0.32$, $k \lesssim 11.5\ {\rm Mpc}^{-1}$ at $z=0.44$) shown in MeerKAT measurements. For greater visualization, lengths smaller than the frequency resolution or the selected k-mode are shaded in gray, where the length is less likely to be resolved by the data.}
 \label{fig:which_one}
\end{figure*}

To address these questions and refine our theoretical framework to better align with measurements, we employ the \textsc{NuetrualUniverseMachine} galaxy-\hi\ catalog, which is built on the top of \textsc{UniverseMachine} simulation \citep{Behroozi2019} and derived using the latest empirical model proposed by \cite{Guo2023}. In the empirical model, \hi\ statistics, e.g., \hi Mass Function (HIMF), \hi stellar mass relation and \hi abundance $\Omega_{\rm HI}$, are well constrained by existing observational measurements (see more details in Section \ref{sec: data}). 

Given that the catalog lacks particle information, and therefore velocity information, it is critical to model and assign velocity profiles to each source. We explore the impact of its shape on the \hi power spectrum by testing three toy models. Then we construct a model for the velocity dispersion, and use some statistics of emission line profile measured by extragalactic \hi surveys to constrain the parameters in our model. The predicted 21cm brightness power spectrum is calculated as Equation \ref{eq:ps_21} in this work.
\begin{eqnarray}
\label{eq:ps_21}
    P_{21\ \rm{cm}}(k,\mu, z)=\overline{T}^2_{\rm b}(z)\cdot P_{\rm HI}(k,\mu,z)
\end{eqnarray}
where $\overline{T}_b$ is the mean brightness temperature; $P_{\rm HI}(k,\mu, z)$ is the cylindrical power spectrum of normalized \hi density fluctuation $\delta_{\rm HI}=\rho(\textbf{x})/\overline{\rho}-1$. $\overline{T}_{\rm b}$ can be determined as follows \citep{Berti2023},
\begin{eqnarray}
    \overline{T}_{\rm b}(z)=180\ \Omega_{\rm HI}(z)h\dfrac{(1+z)^2}{H(z)/H_{\rm 0}}{\rm mK},
\end{eqnarray}
where $\Omega_{\rm HI}$ is the \hi\ abundance, $H_0\equiv100h\ {\rm km\ s^{-1}}{\rm {Mpc}^{-1}}$ is the Hubble constant, $H(z)$ is the Hubble parameter. By normalizing the \hi abundance in $P_{\rm HI}$ and incorporating it into $\overline{T}_{\rm b}$, this work separately presents the \hi bias $b_{\rm HI}$ and \hi\ abundance $\Omega_{\rm HI}$.
Lastly, we illustrate how different observational settings affect the \hi power spectrum and constrain our model based on the measurements.

This paper is organized as follows: Section \ref{sec: data} provides a brief introduction to the simulations and measurements used in this work. Section \ref{sec: emission line profile} describes how we simulate the emission line profile based on \hi\ and subhalo/galaxy catalogs. Section \ref{sec: results} presents the results, including the cylindrical 2D and spherically averaged 1D power spectra in redshift space, \hi\ bias, and a discussion on the shot noise term. Section \ref{sec: obs} examines how observational settings impact the measured power spectrum and the constraint results. Finally, Section \ref{sec: conclusion} summarizes the conclusions of this work. We adopt a flat, $\Lambda$CDM cosmology with parameters ($\Omega_m=0.307, \Omega_{\Lambda}=0.693,h=0.678,\sigma_8 = 0.823, n_s = 0.96$) consistent with \textit{Planck} 15 results \citep{Planck2016}.

\section{Overview of The Data}
\label{sec: data}
In this section, we give a brief introduction to the mock \hi catalogues applied in this work including \textsc{NeutralUniverseMachine} and IllustrisTNG100-1, and also the measurements of the 21cm power spectrum from the MeerKAT telescope. Most of our work is based on \textsc{NeutralUniverseMachine} catalogue, so all results are based on this, unless otherwise specified.
    \subsection{Modeling \hi Catalogs}
        \subsubsection{\textsc{NeutralUniverseMachine} Catalog}
        \label{sec:NUM}
        \textsc{NeutralUniverseMachine} (hereafter ``NUM'') is an empirical model framework proposed by \cite{Guo2023}, which can simultaneously give self-consistent predictions for the evolution of \hi and molecular hydrogen (H$_2$) in a redshift range of $0\leq z\leq 6$. To achieve this, they built the model as functions of redshift $z$ and properties of (sub)halos and galaxies (see their Equations 1 and 15) and jointly constrain the 15 free model parameters using observational measurements in the redshift range of $0<z<6$. The best-fitting model successfully describes the \hi-halo relation, as well as the evolution of the cosmic \hi\ abundance. Explicitly, the \hi mass within a halo or subhalo has been found to be correlated with the virial halo mass $M_{\rm vir}$ \citep{Battye2013, Villaescusa-Navarro2018, Guo2020}, the halo formation time $z_{\rm form}$ \citep{Guo2017, Li2022}, and the star formation rate (hereafter ``SFR'') \citep{Guo2021,Saintonge2022}, which motivates the empirical functional form of the \hi\ gas in the \textsc{NeutralUniverseMachine} model.

        In this paper, we use the public \hi\ catalogues of the \textsc{NeutralUniverseMachine} model, which is derived from the public catalogs of \textsc{UniverseMachine} DR1. These catalogs are generated by running the empirical model \textsc{UniverseMachine} \citep[see][]{Behroozi2019} in the Bolshoi-Planck N-body simulation \citep{Klypin2016}, with a box size of $250\ {\rm Mpc}/h$ and a dark mass particle resolution of $2.3\times10^8\ \msun$.

        \subsubsection{IllustrisTNG100-1 Catalog}
        The IllustrisTNG (hereafter ``TNG'') simulation (\citealt{Marinacci2018,Naiman2018,Nelson2018,Pillepich2018,Springel2018, Nelson2019}) is a set of state-of-the-art hydrodynamical simulations, containing a wide range of physical processes of galaxy formation. In this work, we adopt the simulation set of TNG100-1 with a box size of $110.7$~Mpc and a dark matter particle mass resolution of $7.5\times10^6\msun$.

         To be more accurate, we employ the updated \hi mass calculated by \cite{Diemer2018}. In previous works, the ultraviolet photons emitted from quiescent gas cells with density below a so-called star-formation threshold are usually ignored \citep{Marinacci2017} or set to the cosmic UV background \citep{Lagos2015} while \cite{Diemer2018} consider the contribution from these cells to calculate the atomic-to-molecular transition in TNG simulation. There are several models discussed and updated in this paper, and for simplicity, we employ the ``K13" model, which is an analytical model to predict the fraction of molecular Hydrogen $H_2$ and \hi on the top of neutral gas surface density and metallicity \citep{Krumholz2013}.

    \subsection{Auto-Correlation \hi Intensity Mapping Measurement}
    \label{sec: meerkat_data}
    \cite{Paul2023} (hereafter \citetalias{Paul2023}) reported the first detection of the non-linear \hi power spectrum independently at redshifts $z=0.32$ and $z=0.44$ using the L-band ($900 - 1670 \rm MHz$) data obtained by the interferometer mode of MeerKAT radio telescope. As the precursor to SKA radio arrays, the MeerKAT radio telescope can be used in two different modes (single-dish mode and interferometer mode) for different cosmology purposes \citep{SKA2020}. MeerKAT consists of 64 $13.5\,$m dish antennas located in South Africa. \footnote{\url{https://skaafrica.atlassian.net/wiki/spaces/ESDKB/overview}}

    The measurements of the \hi power spectrum came from the Deep2 field ($\alpha=04^{\rm{h}}13^{\rm{m}}26.4^{\rm{s}}, \delta=\ang{-80;00;00}$ in the Southern Hemisphere), in which only limited foreground radio sources exist. The visibilities were calibrated via the \texttt{\sc{processMeerKAT}} \citep{Collier2021} software pipeline. They chose two RFI-free frequency ranges centered at $986$ and $1077.5$\,MHz with a bandwidth of ${\sim}\,46$\,MHz, equivalent to $\Delta z \sim 0.06$, small enough to ignore the cosmological evolution. The foreground avoidance technique was applied in the analysis, and the frequency resolution of the data is $209 {\rm kHz}$, equivalent to a comoving distance $0.9596\ \mathrm{Mpc}$ at $z=0.32$ and $1.065\ \mathrm{Mpc}$ at $z=0.44$ (see calculation details in Section \ref{sec: FR}).

\section{Simulated Emission Line Profile}
\label{sec: emission line profile}

\hi emission line profiles have been widely detected by extragalactic \hi surveys in the local Universe ($z\leq0.1$) \citep{Barnes2001,Haynes2018,Maddox2021,Zhang2024}. At higher redshifts, most \hi\ sources are too distant to be identified by current radio telescopes, which falls short of the completeness requirements needed for common statistical studies.

When applying the intensity mapping technique, such selection effects are eliminated, and the velocity dispersions in the combined signals can still be resolved by some interferometer arrays like MeerKAT and therefore impact the measured \hi\ power spectrum.

To simulate the emission line profile, the direct way is to run hydrodynamic simulations which include the velocity information of particle data \citep[e.g.,][]{Villaescusa-Navarro2018, Dave2019}. However, hydrodynamic simulations themselves are typically computationally demanding, and the particle \hi data are usually obtained during the post-processing, making them heavily model-dependent and difficult to parameterize. Another method, which applies mock galaxy-based or halo-based \hi catalog, is more widely adopted, as it is in this work.

Using mock catalogs is more convenient and reasonably reliable, since in the low redshift universe, almost all \hi reside in galaxies and halos \citep{Villaescusa-Navarro2018}. Based on different N-body dark matter simulations, both \cite{Sarkar2018} and \cite{Sarkar2019} assume that \hi resides in halos and has different velocity profiles as functions of the properties of their host dark matter. 

\begin{figure*}
    	\centering
    	\includegraphics[width=1.0\textwidth]{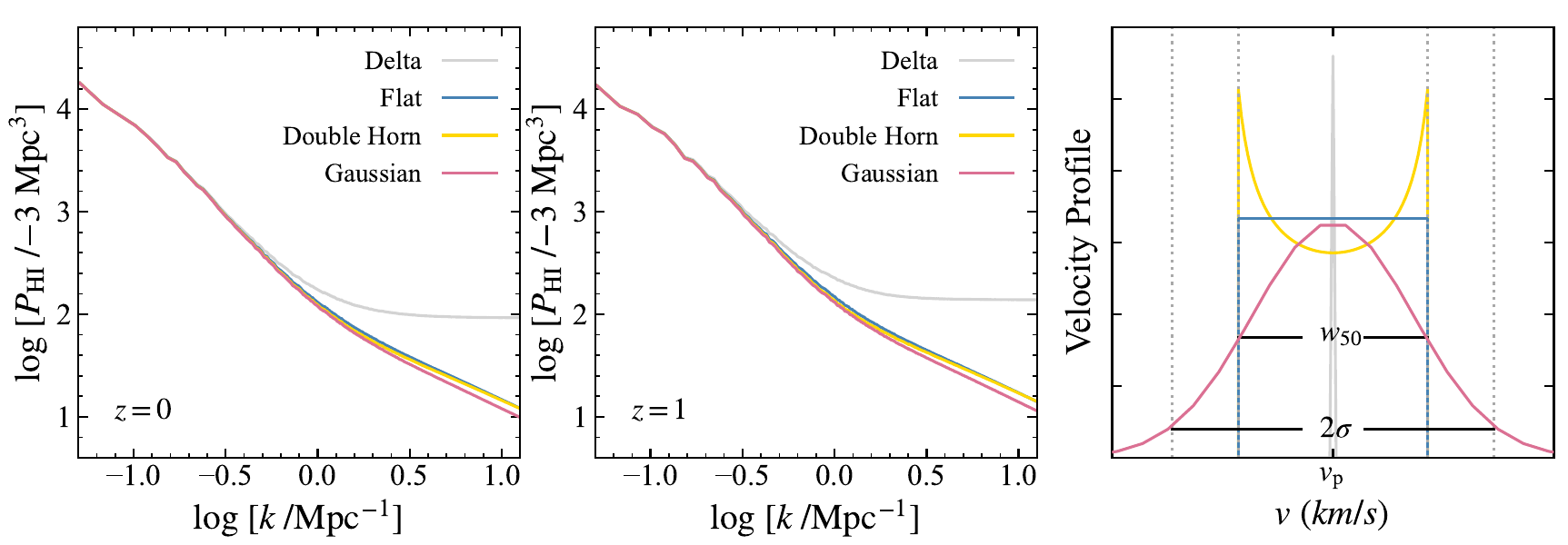}
    	\caption{\hi power spectrum in redshift space with different functions of velocity profiles. As illustrated in the right panel, different colors represent different functions of the profiles: pink for Gaussian distribution, blue for uniform distribution and the yellow for Double Horn distribution. Especially, the grey lines correspond to the case without assigning any profile to the \hi sources and are shown only for comparison. To be general, two redshifts $z=0$ (left) and $z = 1$ (mid) are displayed in the figure.}
    	\label{fig:vf_function}
    \end{figure*}

\cite{Sarkar2019} constructed a double-horn-like model for the velocity profile as a function of halo circular velocity $v_{\rm cir}$. They found that including the inner motion of the \hi disc could considerably enhance the effect of the FoG, that is, suppress the \hi power spectrum in large $k$ modes, and that the suppression can be modeled by a Lorentzian damping profile. \cite{Zhang2020} adopted the relation between \hi velocity dispersion and the host halo mass found by \cite{Villaescusa-Navarro2018} (see Equation 26). They constructed a Gaussian model for the \hi velocity profile as a function of the host halo mass.

Nevertheless, the plausibility of using halo-based models without considering subhalo/galaxy structures to predict \hi\ power spectra needs to be deliberated, especially on non-linear scales, and the shape and width of \hi emission line profile are modeled quite differently in these works. In this section, we verify that both galactic \hi catalogs and an accurate velocity profile are needed to model the real signal and we give a prescription for how to incorporate the velocity profile with the calculation of the \hi power spectrum.

\subsection{Which Elements Matter?}
\label{sec:which one}

To simulate the \hi signal for a certain survey, several factors need to be carefully considered, which include the diameter of \hi discs, stretching length of \hi velocity dispersion in redshift space, the distances between galaxies. We can ask the following questions to elicit the key point. Can the resolution of the \hi survey resolve the structures within halos? Does the velocity dispersion caused by the Doppler effect have a non-negligible effect on the power spectrum? Is the \hi density profile necessary to be modeled? 

Taking the observation by \citetalias{Paul2023} as an example, Figure \ref{fig:which_one} transforms all the factors into length for a better comparison and articulates which elements matter for the observation. Upper panels depict lengths along the LOS direction, accounting for Redshift Space Distortion (RSD) effects, while the lower panels are lengths perpendicular to the LOS direction. The dashed gray lines in all panels are the length resolution,  which is either the frequency resolutions (209\,kHz) along LOS or the largest $k$ modes in the measurements perpendicular to LOS ($k \lesssim 10.5\ {\rm Mpc}^{-1}$ at $z=0.32$, $k \lesssim 11.5\ {\rm Mpc}^{-1}$ at $z=0.44$). 

The right white region of the grey lines represents lengths that can be resolved by the data, while left shadow regions represents length smaller than the resolution. Imagine a mesh field gridded by the resolution lengths within which \hi\ sources are placed. If sources are precisely located at the center of each grid, then all length smaller than the resolution cannot be resolved. However, in reality, many sources are situated at random positions within the grids. Therefore, not all lengths smaller than the resolution are necessarily unresolved by the telescope, as not all sources are exactly at the center of the grid. Therefore, lengths slightly smaller than the resolution are still likely to be resolved. It is tricky to specify a particular length value below which the factor and corresponding lengths can be neglected, so we recommend to disregard factors of lengths much smaller than the resolution and simulate the remaining factors accurately.

In this figure, the black lines illustrate the histogram of diameters of \hi density profile that are estimated using the \hi size-mass scaling relation obtained by \cite{Wang2016}; red lines is the histogram of the simulated emission line width denoted as ``$W_{\rm 50}$'' obtained by fitting to the width function of $W_{\rm 50}$ measured by ALFALFA survey (see more details in section \ref{sec:w50}); blue lines represent the histogram of ``galaxy gap'', which is the distances between galaxies inside the same host halos in redshift or real space. Notably, the blue histograms differ considerably along the LOS and vertically because of the peculiar velocity of galaxies.

There are three points that can be confirmed through Figure \ref{fig:which_one}. Firstly, it is necessary to adopt galaxy-based \hi catalogs, because the telescope can resolve the galactic structures inside halos. As shown in all of the panels, the majority of blue histograms are in the right-hand of the dashed lines. Second, it is unnecessary to consider the \hi density profiles inside the \hi discs. The sizes of \hi discs are much smaller than the scale of our concerns in both directions, so it is safe to neglect the \hi\ density profiles and to regard the \hi discs as point masses in this work. Third, although only a small part of the velocity dispersion along the LOS direction is larger than the frequency resolution, it is essential to account for this dispersion in the computation. This is because the larger velocity dispersion, the larger \hi mass \citep[e.g., see Figure 3 in][]{Oman2022} and thus including the velocity dispersion can severely suppress the shot noise term (see Equation \ref{eq:SN_updated}, where the large mass term accounts for a large fraction).

\subsection{Calculation Methods in Simulations}

We outline the steps we take to generate a \hi density map in redshift space and compute the \hi power spectrum as follows.

    1. Calculate the \hi position in redshifts space and set periodic boundary conditions;
    
    \begin{equation}
    \label{eq:rsd_pos}
        {\bf s}={\bf x}+\dfrac{v_\parallel(1+z)}{H(z)},
    \end{equation}
    where the ${\bf x}$ is the comoving position of \hi\ sources in real space; ${\bf s}$ is the comoving position in redshift space; $v_\parallel$ is the peculiar velocity projected in line of sight (LOS) direction. Plane-parallel approximations are adopted here; $H(z)=H_0\cdot E(z)$ is the Hubble parameter.
    
    2. Evenly cut the $250^3\ {\rm Mpc^3}/h^3$ simulation box into $1200^3$ grids. If frequency resolution is considered, the grid number along this axis should be reset accordingly. Then locate the center of each \hi\ source in the mesh according to their position ${\bf s}$. 

    3. Assign a single-peak or double-horn emission line profile along the line of sight (LOS) for each \hi\ source, with the profile width, such as $W_{\rm 50}$ or $W_{\rm 20}$ in extragalactic surveys, being solely determined. In this work, we use a Gaussian function for the shape and Full Width at Half Maxima (FWHM) for the width, also known as $W_{\rm 50}$ (see more details in Section \ref{sec:w50}). The standard deviation $\sigma$ of the Gaussian profile can be calculated accordingly.
    \begin{equation}
        \sigma_i = \dfrac{W_{{\rm 50}, i}}{2\sqrt{2\ln{2}}}
    \label{eq: sigma}
    \end{equation}
    Along the chosen LOS direction, select the grids within the $\pm 2\sigma$ length near the $i$th source and assign its mass $M_i$ to the grids according to the error function.
    \begin{equation}
        M_{\rm n}= M_i\int^{z(n+1)}_{z(n)} N(z_{\rm s},\sigma_{s}) {\rm d}z',
    \label{eq:mass ratio}
    \end{equation}
    where the $M_{\rm n}$ is the mass assigned to $n$th grid; $z(n)$ is the position of left side of the grid, and $z(n+1)$ is the right side; $N(z_{\rm s},\sigma_{s})$ is the normalized Gaussian distribution.

    We only consider the $\pm 2\sigma$ range of the Gaussian profile, which provides a sufficiently large span to accurately compute the power spectrum. In cases where both the start and end points of the profile reside within the same grid, the Nearest Grid Point (NGP) algorithm is employed to assign the source to the nearest grid regardless of the profiles.

    4. Normalize the density field to a mean-centred fluctuation field; $\delta(\textbf{s})=(\rho(\textbf{s})/\ \overline{\rho})-1$
    
    5. Use the Fast Fourier Transform (FFT) to calculate the 3D power spectrum and then spherically average the 3D power spectrum into wavebands of $k$ to give the 1D power spectrum, and calculate the amplitude according to Equation \ref{eq:ps_21}. If observational settings are considered, then implement the cuts of $k$ modes in the 3D power spectrum.

    6. Multiply the power spectrum of \hi\ density fluctuations with the square of the mean brightness temperature $\overline{T}^2_b$.

    We assign each stretched \hi source in our particle-mesh code following the Nearest Grid Point (NGP) methods. Notably, we do not adopt the Cloud-in-Cell (CIC) or other high-order smoothing methods (hereafter referred to as "CIC methods") and only adopt NGP due to several factors. Firstly, as long as the sampling frequency is much larger than the frequency of our concern, NGP can be accurate enough. Secondly, as some low-pass filters are narrower than NGP \citep[e.g., see Figure 8 in][]{Cunnington2024}, CIC methods damp the power spectra at small scales and also distort the shot noise plateau if not properly deconvolved. Furthermore, Fourier aliasing effects become too pronounced if the CIC or other kernels are deconvolved without performing the interlacing method to remove the aliasing effects. Hence, although the NGP methods can slightly suppress the power spectra in the smallest scales, given the limited computation cost, we believe that NGP is the optimal compromise.

\subsection{Emission Line Profile Model}
\label{sec: emission line model}

    \begin{figure}
        \centering
        \includegraphics[width=0.47\textwidth]{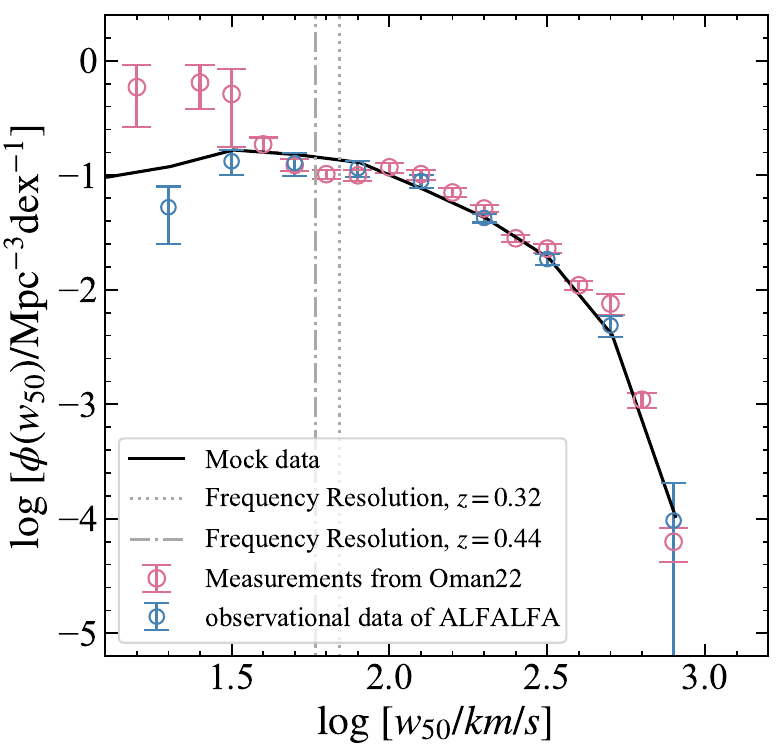}
        \caption{Width Function of $W_{\rm 50}$ obtained from ALFALFA (circles with error bars) and mock data (black solid line). Pink circles with error bars are the results of \citet{Oman2022}, and blue circles are obtained in this work from the data from ALFALFA. The black lines are constructed by $V_{\rm max}$ in the NUM catalog data. For reference, grey dotted and dashed lines are the frequency resolution of MeerKAT at $z=0.32$ and $z=0.44$, respectively.}
        \label{fig:w50}
    \end{figure}

The emission line profiles are generally determined both by the velocity profile and the inclination angle $i$ between the LOS direction and the normal to the galaxy disc. We consider two factors of the velocity profile: 1) the shape of the profile and 2) the profile width. We verify that the shape of the profile has a subtle impact on the power spectrum, while the width plays a key role.
    \subsubsection{Shape of Emission Line Profile}
    There are two common profile shapes which correspond to two different kinematic scenarios. 1) Double Horn, found in spiral galaxies where \hi is usually distributed in the outskirts of galaxies and is in regular rotation motions in the discs. Typically, these sources have a relatively large inclination angle, close to edge-on; 2) Gaussian-like distribution, found in the elliptical galaxies where the random motions in the discs dominate or in face-on \hi discs. Both shapes are observed in extragalactic \hi surveys \citep[e.g., see Figure 4 in][]{Masters2019}. Beyond these two, there are also some other irregular shapes that cannot be well defined. To explore the impact of the shape of velocity profiles on the \hi power spectrum, we use three toy models corresponding to three shapes, Gaussian, double-horn and flat as seen in the right panel of Figure \ref{fig:vf_function}.
    \begin{figure*}
    	\centering
    	\includegraphics[width=\textwidth]{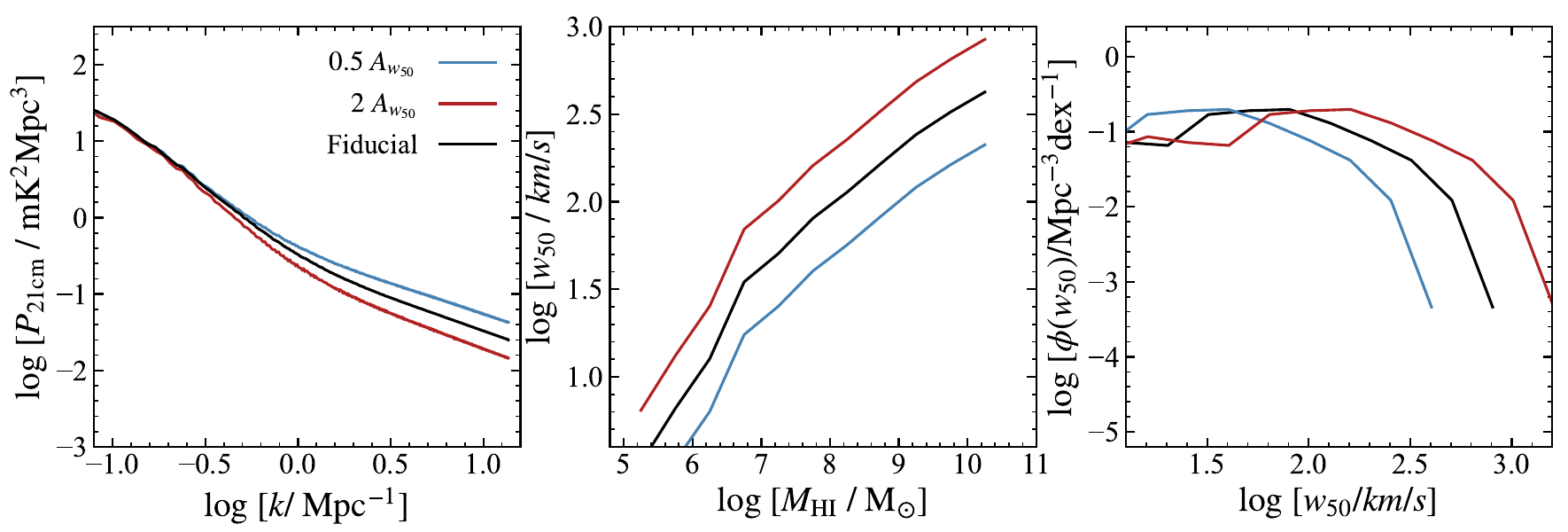}
    	\caption{Impacts of parameter $A_{W_{\rm 50}}$ in the $W_{\rm 50}$ model on the 21cm power spectrum. The left, middle, and right panels display the power spectra, median of the $W_{\rm 50}$-\hi relation, and the width function of $W_{\rm 50}$, respectively. Blue lines represent the statistics with one parameter halved, while red lines depict those with parameters doubled.}
    	\label{fig:ps_w50}
    \end{figure*}
    Shown as a schematic, the coloured lines in the right panel provide insight into how these shapes look like. The colours of the spectra lines in the left and middle panels correspond to the shapes depicted in matching colors in the right panel: pink for the Gaussian distribution, blue for uniform distribution, and yellow for double-horn distribution. The double-horn function is the same as the fiducial model of \cite{Sarkar2019} (see Equation 5 in this paper); the $\sigma$ parameter in the Gaussian function is determined by the assigned $W_{\rm 50}$.

    The left and middle panels of Figure \ref{fig:vf_function} illustrate how different shapes affect \hi power spectrum at two example redshifts, $z=0$ (left panel) and $z=1$ (mid panel). For a fair comparison, each \hi source is assigned a fixed profile width, denoted as $W_{\rm 50}$, i.e., the Full Width at Half Maxima (FWHM), determined by an empirical model (see more details of this part in Section \ref{sec:w50}). The mass of each \hi source is equal to the integral of the profile, which means that the amplitudes of the profiles are proportional to the corresponding \hi mass.
    
    The differences between the coloured lines in the left and middle panels are subtle, which means that the impact of the profile shape on the \hi power spectrum can be neglected. We also show the power spectrum of \hi sources without the velocity profile as grey lines in all panels. The large differences between the coloured lines and the grey lines emphasise the importance of including the profile width $W_{\rm 50}$ for small scales. For large scales with $k < 0.5\ {\rm Mpc^{-1}}$, the power spectra are not affected at these redshifts.

    \subsubsection{Emission Line Width $W_{\rm 50}$}
    \label{sec:w50}
    \begin{figure*}
	\centering
	\includegraphics[width=\textwidth]{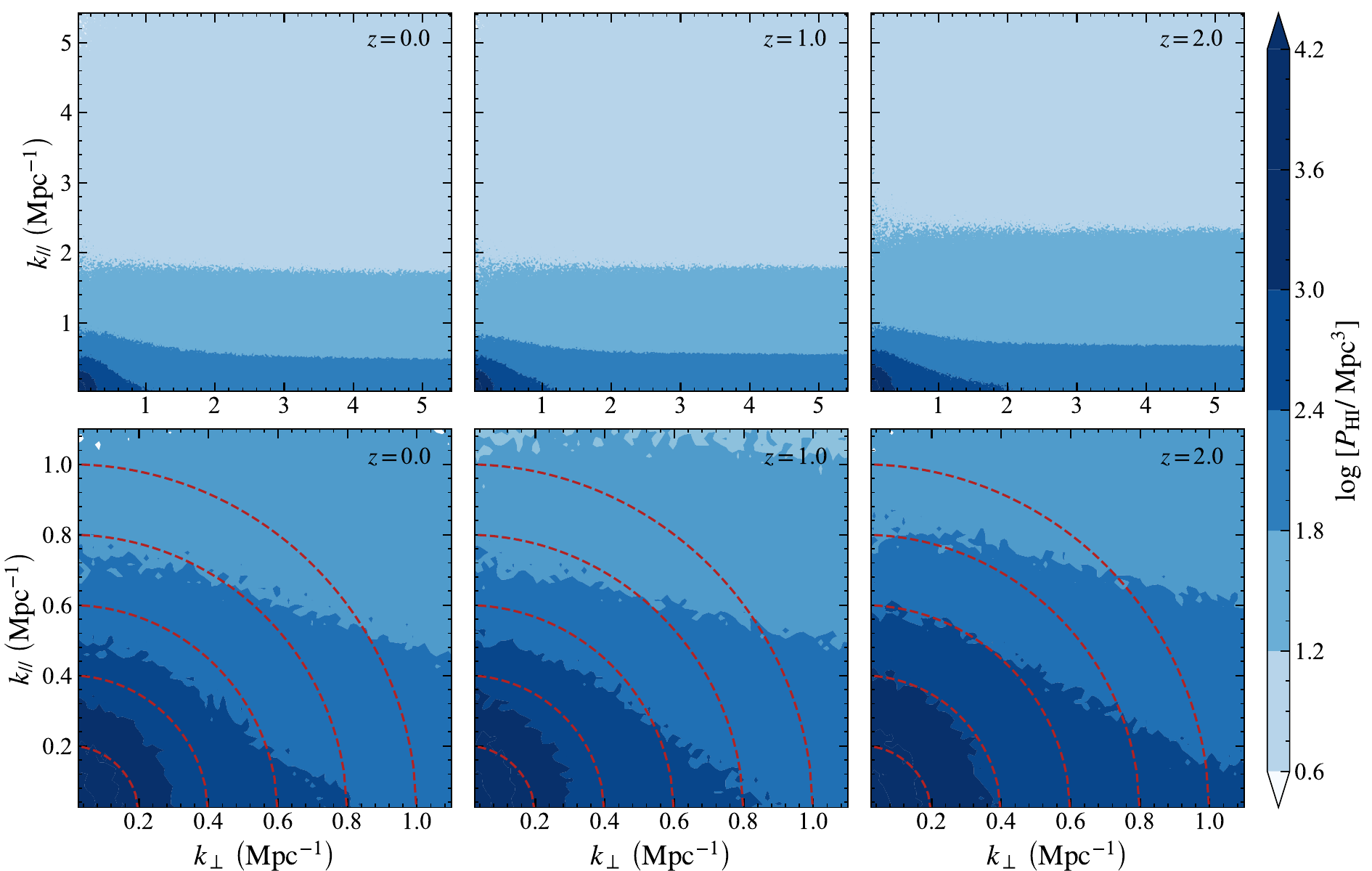}
	\caption{From left to right, this figure display the 2D Power Spectrum of 21cm signal emitted by \hi, $P_{\rm 21cm}(k_\parallel,k_{\perp})$, in redshift space at various redshifts ($z=0.0, 1.0, 2.0$). The bottom panels are the zoom-in of the upper panels. The FoG effect is evident in the upper panel with a large range of k-modes, which exhibits a step-like pattern, while the Kaiser effect induced by large-scale compression is very pronounced in the lower panel with a small range of k-modes.}
	\label{fig:2dps}
    \end{figure*}
    $W_{\rm 50}$, also known as Full Width at Half Maxima (FWHM), is a common measure used to describe the width of emission line profiles in extragalactic \hi\ observations \citep[e.g., see][]{Barnes2001,Zavala2009, Zwaan2010,Haynes2018, Oman2022,Zhang2024}. By comparing the grey lines and the coloured lines in Figure \ref{fig:vf_function}, it can be inferred that the value of $W_{\rm 50}$ significantly influences the \hi power spectrum at small scales. Since $W_{\rm 50}$ is not available in the NUM catalogue, we have to quantitatively determine the $W_{\rm 50}$ for each \hi disc. 

    According to the Newtonian dynamics, the cold gas skirts surrounding spiral galaxies should share the same gravity potential with their resident collisionless cold dark matter halo. Therefore, it is commonly assumed that the velocities measured in the flat part of the rotation curves $v_c$ of spiral galaxies and the surrounding gas are equal to the maximum velocity $V_{\rm max}$ of their resident (sub)halos \citep[e.g., see][]{Gonzalez2000, Zavala2009, Chauhan2019}. For these \hi-rich galaxies, the emission line width $W_{\rm 50}$ can be approximated as
    \begin{eqnarray}
        W_{\rm 50} =  2\cdot V_{\rm max}\sin{i}\ ,
    \end{eqnarray}
    where $i$ is the random inclination angle between our LOS and the norm of disc, assumed to be entirely random between $0$ and $\pi$, not perfect, but economical and stable. Although elliptical galaxies introduce additional complexity, this work ignores the difference between them and spiral galaxies due to their limited \hi\ content.
    
    Under this assumption, it is supposed that the velocity function, representing the abundance of galaxies as a function of their circular velocity $v_c$, follows the prediction from the Cold Dark Matter (CDM) model \citep[e.g.,][]{Cole1989, Gonzalez2000, McGaugh2012}. However, discrepancies have arisen in the low-velocity end. Measurements from the ALFALFA survey suggest that there is a serious deficiency for galaxies at the low-velocity end compared to what is expected from CDM simulations \citep{Zavala2009, Klypin2015}. Several hypotheses have been proposed to address this problem. Firstly, the warm dark matter can suppress the abundance of small dark matter halos since it is more difficult for small halos to captures WDM particles with larger velocity and smaller mass compared with CDM particles \citep{Bode2001, Zavala2009, Klypin2015}; Secondly, the limited SNRs and the selection effects in the \hi\ extragalactic surveys are also considered as ways out \citep[e.g., ][]{Chauhan2019, Oman2022, Sardone2024}; Thirdly, it is also possible that for low-velocity galaxies/\hi sources, the equal velocity assumption might fail, i.e., $v_c \neq v_{\rm max}$ due to various baryonic effects \citep[e.g., see ][]{Maccio2016,Brooks2017,Verbeke2017}. Given the ongoing debates surrounding the VF function, this work adopts the $v_c \neq v_{\rm max}$ assumption and expresses $W_{\rm 50}$ as a double power law function of $v_{\rm max}$.

    \begin{eqnarray}
        W_{\rm 50} = \dfrac{A_{W_{\rm 50}}\cdot V_{\rm max}\sin{i}}{\left(V_{\rm max}/V_0\right)^{-\alpha} + \left(V_{\rm max}/V_0\right)^{\beta}}
    \label{eq:w50},
    \end{eqnarray}
    where $i$ is the random inclination angle between our LOS and the norm of disc, and ${A_{W_{\rm 50}}, V_0, \alpha, \beta}$ are the free parameters in this empirical model.

    Figure \ref{fig:w50} displays the width function of $W_{\rm 50}$ obtained from ALFALFA surveys (pink and blue circles with error bars) and our mock data at $z=0$ (solid black lines). The blue circles are the measurements obtained from ALFALFA in this work and we use them to constrain the formula, while the pink circles are the measurements obtained by \cite{Oman2022} shown here for reference. The gray dotted line and the dashed dotted line represent the frequency resolution of MeerKAT shown in Figure \ref{fig:which_one}.
    
    We use the width function of $W_{\rm 50}$ measured by the ALFALAF survey to constrain the parameters and find the best-fit values as $A_{W_{\rm 50}}=4.868$, $V_0=76.053$ \kms, $\alpha=2.075$, $\beta=0.675$. The ALFALFA survey focuses primarily on the local universe ($z\leq 0.05$), thereby lacking constraints on the width function at higher redshifts. However, we assume that $W_{\rm 50}$ only evolves as $V_{\rm max}$, which is considered properly in the simulation. We employ the best-fit parameters as our fiducial model to estimate $W_{\rm 50}$ at all low redshifts ($0\leq z\leq 1$).

    We further investigate how variations in parameters within our fiducial model influence the \hi power spectrum. From left to right panels, Figure \ref{fig:ps_w50} illustrates how statistics including 21cm power spectra, median of $W_{\rm 50}$-$M_{\rm HI}$ relation and width function of $W_{\rm 50}$ change with parameters in Eq. \ref{eq:w50}. The black lines in all panels of Figure \ref{fig:ps_w50} are our fiducial model, shown as a reference, while the blue/red lines represent parameters being halved/doubled. All four parameters are tested, but only $A_{W_{\rm 50}}$ is presented because the \hi power spectrum is insensitive to all parameters except $A_{W_{\rm 50}}$. The specific reason for this phenomenon is explained in further detail in Section \ref{sec: model}.

\begin{figure*}
	\centering
	\includegraphics[width=\textwidth]{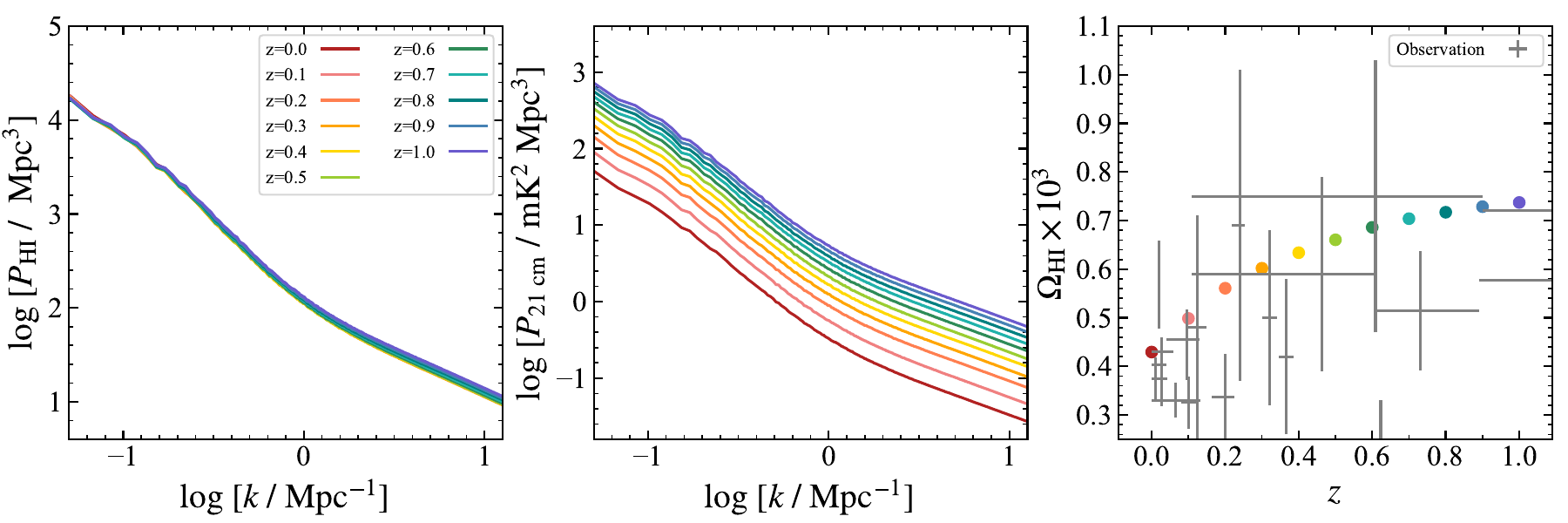}
	\caption{Angular averaging 1D power spectrum of \hi\ fluctuation in redshift space (left panel) and the corresponding 21cm signal (mid panel). Right panel display the $\Omega_{\rm HI}$. Different colors represent mock powers pectrum or \hi abundance at different redshifts ranging from 0.0 to 1.0. The grey error bars in the right panel represents the measurements of cosmic \hi abundance.}
	\label{fig:1dps}
\end{figure*}

\section{Results}
\label{sec: results}

In the analysis of this section, the cylindrical power spectrum, 1D power spectrum, theoretical shot noise term and scale-dependent \hi bias are presented without incorporating any additional observational effects. All calculations are implemented in the redshift space except for the \hi bias.
\subsection{2D Power Spectrum}
\label{sec:2DPS}

In Figure \ref{fig:2dps}, we present the cylindrical power spectra, referred to as 2D power spectra (2DPS), of the \hi density fluctuation field. The bottom panels as the zoom-in of the upper panels illustrate the Kaiser effect, a gravity-driven compression of the structure on large scales in redshift space, which manifests itself as elongation along the parallel direction in Fourier space.

The upper panels depict the FoG effect, prominently observed in large $k$ modes. It is worth noting that the FoG effect is more of a plane that descends along the parallel direction, not exactly a step-like pattern shown in this contour figure due to the discrete colour bar. This pattern arises due to two primary factors. Firstly, in the perpendicular direction, the small diameters of galactic \hi sources allow the \hi sources to be effectively approximated as point masses, leading to Poisson noise, also known as shot noise. Secondly, the emission line profile of \hi results in a gradual decline along the parallel direction. These combined effects produce the observed step-like pattern at large $k$ modes, which would otherwise resemble a shot noise plateau. 

Our predictions for the 2D power spectrum, particularly the FoG effect on small scales ($k>1\ {\rm Mpc^{-1}}$), closely resemble the results obtained from the hydrodynamical simulations of TNG \citep[see top two rows of Figure 22 in][]{Villaescusa-Navarro2018}.

\subsection{1D Power Spectrum}
\label{sec:1dps}

Figure \ref{fig:1dps} illustrates the evolution of the spherically averaged 1D power spectrum of \hi density fluctuations (left panel) and 21cm signals (mid panel) at different redshifts. The power spectra of \hi density fluctuations and 21cm signal are the $P_{\rm HI}$ and $P_{\rm 21cm}$ in Eq. \ref{eq:ps_21}, respectively.

Different colours denote redshifts ranging from $0.0$ to $1.0$. The power spectra of \hi density fluctuation, denoted as $P_{\rm HI}$ in the left panel, are very similar at different redshifts. However, it is worth emphasising that all spectra are displayed on a logarithmic scale, and therefore evolution regarding \hi\ fluctuation spectrum is present, although small (see more details in Section \ref{sec: hi bias}). A more pronounced decreasing trend is displayed in the middle panel. This trend, as also observed in the measurements in \citetalias{Paul2023}, can be attributed to the decline in abundance of \hi over time. The amplitude of 1D power spectrum of 21cm signal is positively correlated with the \hi abundance $\Omega_{\rm HI}$ as shown by Equation \ref{eq:ps_21}. The decrease occurs because of physical processes such as consumption by star-forming and ionisation by ultraviolet photons. As shown in the right panel of Figure \ref{fig:1dps}, the mock \hi\ abundances $\Omega_{\rm HI}$ evolving with time are shown as coloured points and the measurements of $\Omega_{\rm HI}$ are shown as grey dots with error bars. The measurements are obtained from the extended \hi Mass Function (HIMF) at local universe obtained by the HIPASS survey \citep{Zwaan2005}, the ALFALFA survey \citep{Martin2010} and the ALFA Ultra Deep Survey (AUDS) \citep{Freudling2011, Hoppmann2015}; stacking methods for the $z\lesssim 0.6$ \citep{Lah2007, Delhaize2013,Rhee2013, Rhee2016} and the Damped Lyman$\alpha$ (DLA) systems that arise from the background quasar absorption lines at higher redshifts \citep{Rao2006,Neeleman2016,Rao2017}. More information about \hi\ abundance are listed in the Figure 14 of \cite{Rhee2018}. The coloured mock dots are generally within the errors of the observed ones. 

\begin{figure*}
	\centering
	\includegraphics[width=\textwidth]{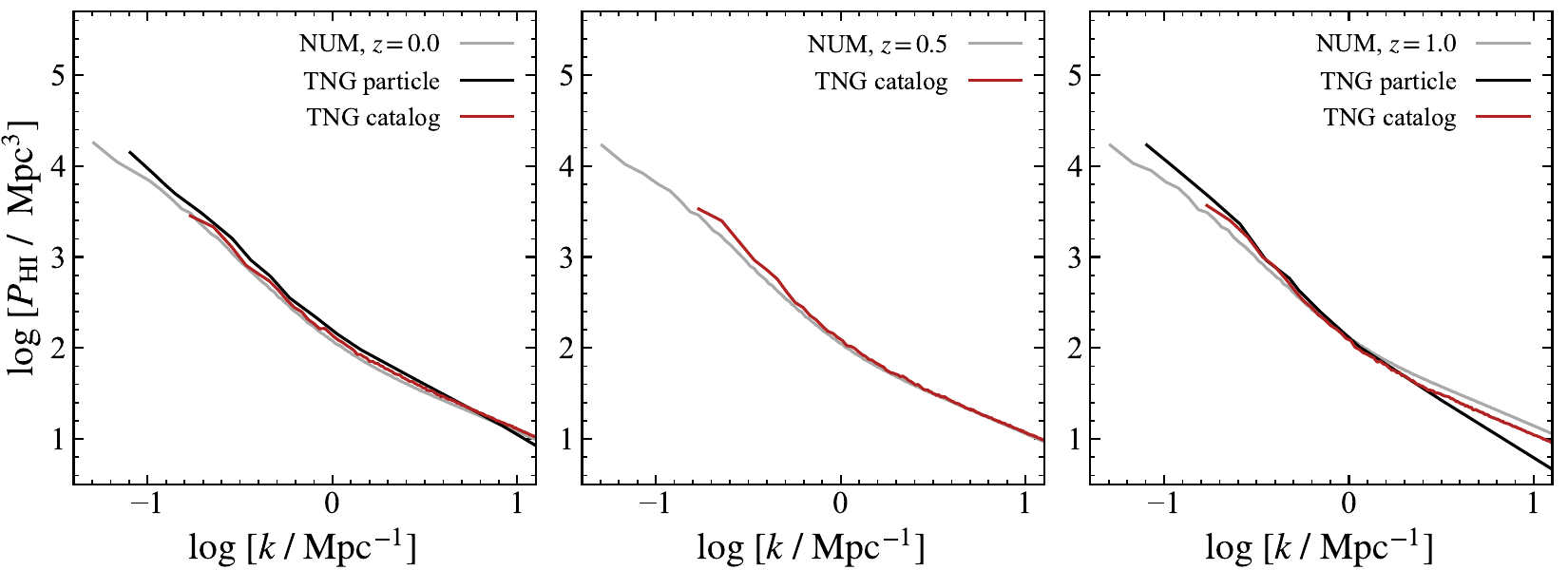}
	\caption{Comparison between our results and the results obtained by TNG at $z=0,0.5,1.0$. The grey lines are our predictions, red lines display the results obtained by applying the same method to the catalog data of TNG simulation, and the black lines are the power spectrum of particle \hi data shown in Figure 21 in \citet{Villaescusa-Navarro2018}. The differences between the red and grey lines are neglectable, and the differences between the particle results and our results are also small.}
	\label{fig:ps_tng_comp}
\end{figure*}

For comparison, we apply our method to the TNG100-1 simulation, with a box size of $75\ {\rm Mpc}/h\ \approx 110\ {\rm Mpc}$, and present the corresponding predictions for the \hi power spectrum in Figure \ref{fig:ps_tng_comp}. We use the TNG \hi catalogs and apply our fiducial $W_{\rm 50}$ model to subhalo $v_{\rm max}$ catalog data in TNG to determine the profile width of each \hi source. The Gaussian shape is adopted for the emission line profiles. Red lines in Figure \ref{fig:ps_tng_comp} depict the outcomes derived from the TNG galaxy catalog, and the grey lines represent our results based on the NUM catalog for comparison. Differences between these lines are negligible, especially on small scales, indicating that our method and predictions are robust across different models and simulations. The only caveat is that the red lines are slightly higher than the grey lines because the linear \hi\ bias of TNG simulation is 20\% larger than that in NUM (see more details in Section \ref{sec: hi bias}). This phenomenon validates the idea that constraints from the shot noise is likely to break the degeneracy between $\Omega_{\rm HI}$ and $b_{\rm HI}$ \citep{Chen2021}.

To validate the precision of our findings, we also conduct a comparison with the results obtained from the particle data from the TNG100-1 simulation \citep[see Figure 21 in][]{Villaescusa-Navarro2018}. The black lines on the left and right panels display the power spectra of the \hi particle data. Only the results of two snapshots are presented, as the power spectra provided by \cite{Villaescusa-Navarro2018} are limited to integer redshifts. The differences between the \hi power spectra of the TNG particles and our results (NUM) are found to be minimal for the local universe, $z=0$. This close agreement reinforces the accuracy of our model and its ability to effectively replicate the particle results.

However, at higher redshifts ($z=1$), discrepancies emerge, although they are not as pronounced. These discrepancies are mainly due to the larger \hi bias of the TNG simulation on large scales and the presence of either sizeable \hi discs or sparse but non-negligible \hi gas distributed outside galaxies on small scales.

\subsection{\hi Bias}
\label{sec: hi bias}

\begin{figure}
        \centering
        \includegraphics[width=0.47\textwidth]{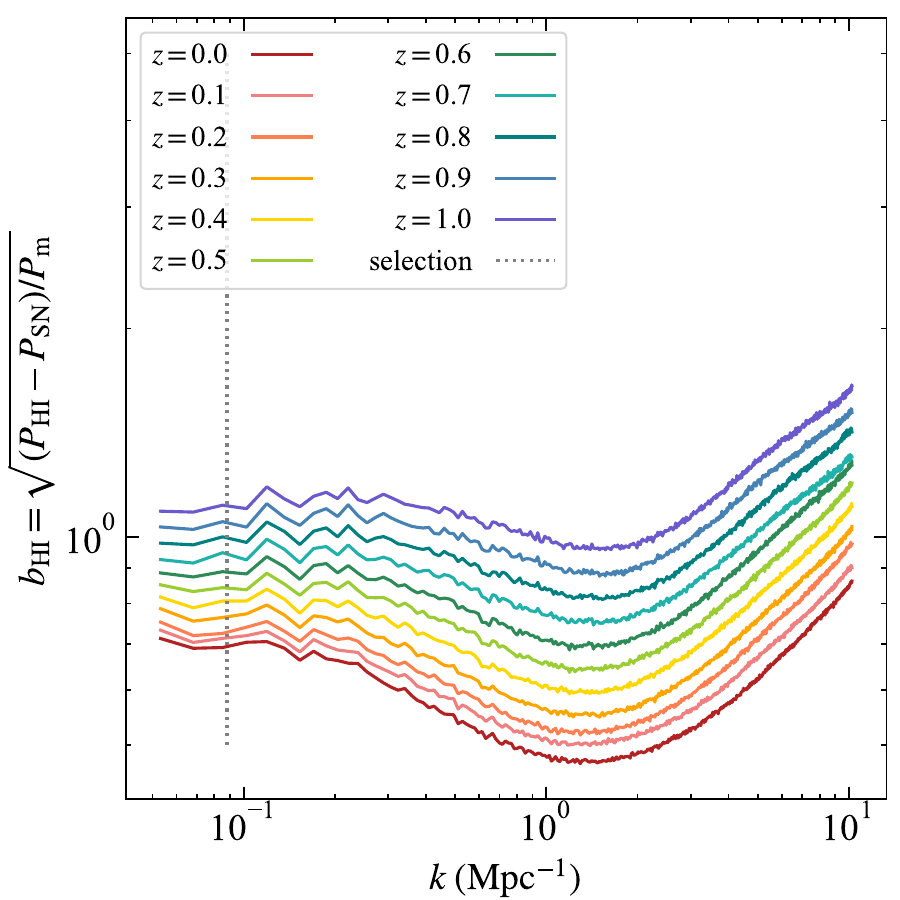}
        \caption{The dependence of \hi bias on scale. Different colors represent different redshifts ranging from 0.0 to 1.0. The bias is calculated by dividing the \hi power spectra by the theoretical matter power spectra and then taking the square root, as indicated by the y-axis label. The shapes of the lines are similar, with amplitudes increasing as redshifts increase. The gray dotted line denotes the $k$ modes below which the linear bias is determined.}
        \label{fig:hi bias}
    \end{figure}

Figure \ref{fig:hi bias} illustrates the dependence of \hi bias on scales, given by $b_{\rm HI}(k)=\sqrt{(P_{\rm HI}-P_{\rm SN})/P_{\rm m}}$, where $P_{\rm SN}$ is the shot noise term; $P_{\rm m}$ is the theoretical non-linear matter power spectrum calculated using CAMB \citep{Lewis2000} assuming the same cosmology as the simulation \citep{Klypin2016, Behroozi2019, Guo2023}. Each colour corresponds to a different redshift, ranging from 0.0 to 1.0. The grey dotted line marks the $k\approx0.1\ \rm Mpc^{-1}$ threshold below which the linear \hi bias is determined \citep{Wang2021}. Specific values of linear bias at different redshifts are presented in Table \ref{table: hi bias}. Due to the limitation of the box size, the bias with $k$ modes less than $0.03\ h/\rm Mpc$ are excluded from the calculation and are not depicted in the figure. The average of $k$ modes less than $0.045\ h/\rm Mpc$ is calculated separately in this work, rather than using the centre of the $k$ bins. This is because the sparsity of $k$ modes near the scale of the simulation box can cause the true average $k$ modes to deviate from the $k$ bin center according to our tests.

Above $k\approx0.1\ \rm Mpc^{-1}$, there is a dip feature that appears at $k\approx1 \rm Mpc$, as reported by \cite{Villaescusa-Navarro2018} and \cite{Spinelli2020} as well. This dip is possibly a result of the bump feature appearing in the non-linear matter power spectrum at this scale. This would also explain the less pronounced dip feature at higher redshift (z>1) as seen in \cite{Villaescusa-Navarro2018, Spinelli2020}, since non-linear effects are less dominant at higher redshifts.

It should be noted that our \hi bias estimate is consistently lower than that provided by other simulations \citep[see, e.g.,][]{Villaescusa-Navarro2018, Spinelli2020}. In contrast, our \hi bias for the local universe is in line with the measurements of ALFALFA reported by \cite{Li2022}, resulting in a 20\% lower estimate. 

A more realistic mock has been generated by \cite{Guo2023} to assess the robustness of the \hi clustering measurements obtained from ALFALFA (refer to the right panel of Figure 6 in their paper). Unfortunately, the mock results suggest that the observation volume of the ALFALFA survey is insufficient to yield robust projected \hi clustering measurements. Fortunately, the \hi catalogue of the local universe by Five-hundred Aperture Spherical Telescope (FAST) is imminent \citep{Zhang2024}. In the near future, a more accurate estimate for \hi bias can be achieved by analysing the combination of ALFALFA surveys and the FAST all sky HI survey(FASHI), providing insights into \hi bias, particularly within the local universe.

\begin{table*}
    \caption{Values of \hi Bias of our simulation at different redshifts}
    \begin{tabular}{lccccccccccc}
    \toprule
         redshift & 0.0 & 0.1 & 0.2 & 0.3 & 0.4 &0.5& 0.6 & 0.7& 0.8 & 0.9 & 1.0\\
    \midrule
    $b_{\rm HI}$ & 0.698& 0.716 &0.732&0.768&0.805&0.843&0.883&0.930&0.984&1.04&1.10 \\
    \bottomrule
    \end{tabular}
    \label{table: hi bias}
\end{table*}

\subsection{\hi\ Shot Noise}
\label{sec: model}
\subsubsection{Theoretical Shot Noise \& \hi Power Spectrum without Shot Noise}
\begin{figure*}
    \centering
    \includegraphics[width=1.0\textwidth]{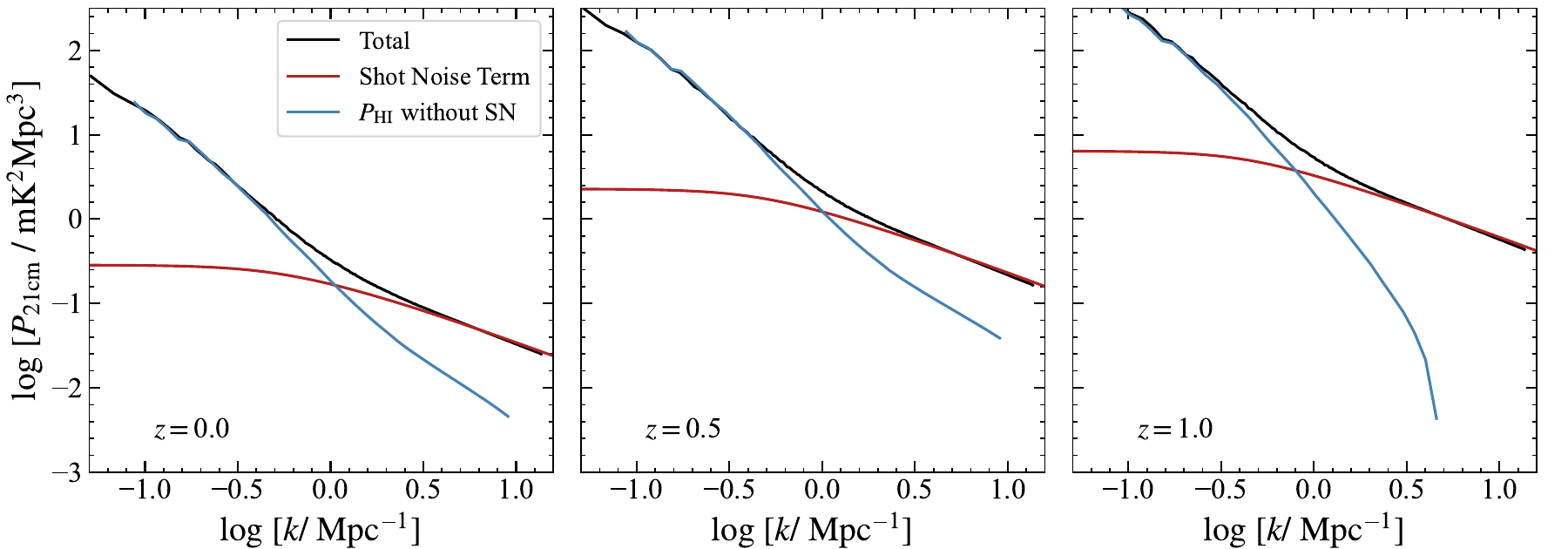}
	\caption{Simulated \hi power spectrum on the top of mock galaxy catalog and theoretical shot noise term. The black lines are the 21cm power spectrum at $z=0.0, 0.5, 1.0$. The red lines are the theoretical shot noise term. Blue lines are the simulated corrected power spectrum calculated by subtracting shot noise term (red lines) from the total power spectrum (black lines).}
    \label{fig:component}
\end{figure*}

Considering the Redshift Space Distortion (RSD) in our model, the position of each source in the redshift space is affected by the peculiar velocity, while it is also stretched due to their inner motions, as discussed in Section \ref{sec:w50}. 
Theoretically, the total power spectrum in the redshift space can be separated into the power spectrum without shot noise $P^{(s)}$ and the shot noise term $P_{\rm SN}$. For each part, we need to consider different RSD effects. Figure \ref{fig:component} displays the total power spectrum and the divided two parts. The black lines are the total power spectrum obtained from our mock data, identical to the coloured lines in Figure \ref{fig:1dps} at the corresponding redshifts.

\begin{figure*}
    \centering
    \includegraphics[width=1.0\textwidth]{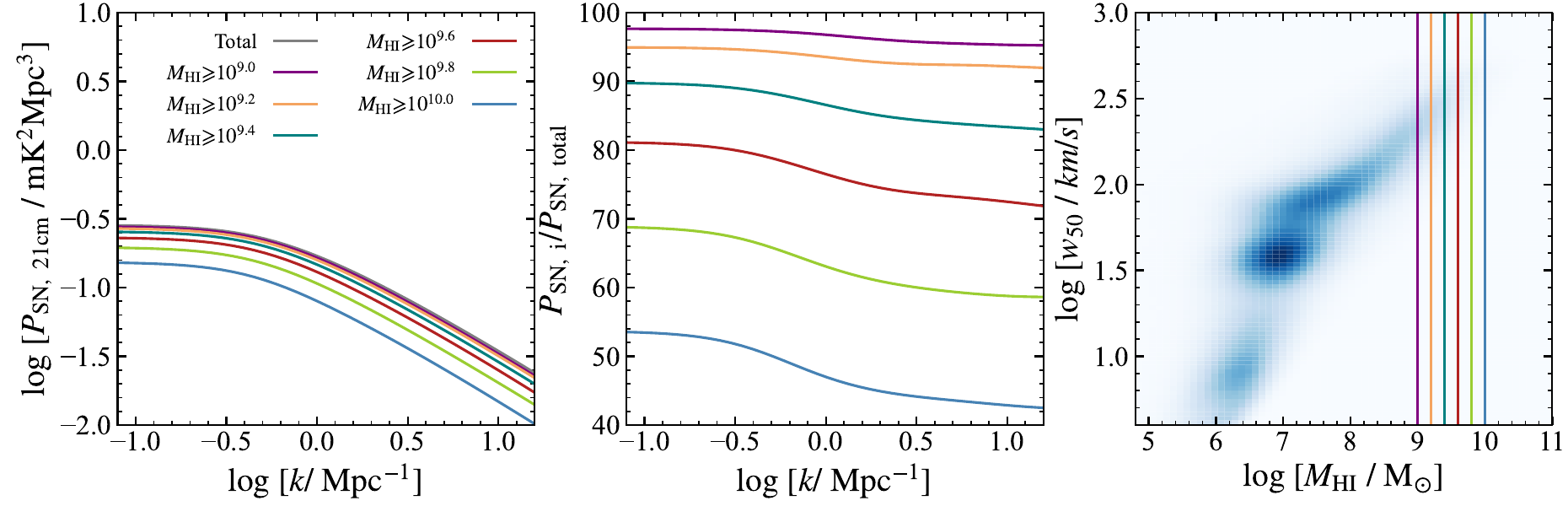}
	\caption{Impact of lower limits in \hi mass function (\hi MF) on the theoretical shot noise term of 21cm power spectrum. Different colors of lines represents different lower limits, as shown by the right panel. The left panel presents the theoretical shot noise with different cuts in \hi MF. Mid panel show the fraction of shot noise with different lower limit in total shot noise. Right panel presents the $W_{\rm 50}$-\hi mass relation in our mock catalog with the colored lines as the lower limits.}
    \label{fig:sn_frac}
\end{figure*}

In the following section, we describe how the shot noise term can be calculated theoretically, as displayed in the red lines of Figure \ref{fig:component}. For this term, the positional information is completely unnecessary; only the \hi mass distribution and the emission line profile of each source are required. Following the assumption in Section \ref{sec:w50}, we replace the Dirac delta function with Gaussian functions. Consequently, we can write the density fluctuation field as follows.
\begin{equation}
\delta(\vec{x})=\dfrac{\rho(\vec{x})}{\overline{\rho}} - 1=\dfrac{V}{M_{\rm t}}\cdot \sum_{i} M_{i} {\rm \delta^{D}}(x-x_i){\rm \delta^{D}}(y-y_i){\mathcal N}(z_i, \sigma_i)-1,
\label{eq:delta_x}
\end{equation}
where ${\rm \delta^{D}}$ is the Dirac function to describe the point sources and $\delta(\Vec{x})$ is the normalized fluctuation or the overdensity; $M_{\rm t}$ is the total mass of all \hi sources $M_{\rm t}=\sum_i M_i$; $V$ is the volume of the simulation box. We set the $z$ axis as the LOS direction and ${\mathcal N}(z_i, \sigma_i)$ as the Gaussian profile. $\sigma_i$ for each source is determined by the simulated $W_{\rm 50}$, similar to Equation \ref{eq: sigma}, and $\sigma_i$ is computed as
\begin{equation}
    \sigma_i = \dfrac{l_{w50}}{2\sqrt{2\ln{2}}},
\end{equation}
where $l_{w50}$ is the stretching length after translating the $W_{\rm 50}$ with units \kms to comoving distance with units $\rm Mpc$ or ${\rm Mpc}/h$.

For simplicity, we only show the LOS direction, while for the perpendicular direction Dirac delta functions still work properly.
\begin{equation}
\begin{aligned}
    \Tilde{\delta}(k_z)&=\dfrac{V}{M_{\rm t}}\cdot \sum_{i} M_{i}\dfrac{1}{\sqrt{V}}\int _V {\mathcal N}(z_i, \sigma_i) \cdot e^{-{\rm i}2\pi k_z z}\ {\rm d}z \\
    &=\dfrac{\sqrt{V}}{M_{\rm t}}\cdot\sum_{i}\left(M_{i} e^{-\frac{1}{2}\sigma_i^2 k_z^2}\right) \cdot e^{-{\rm i}2\pi k_z z_i}
\end{aligned}
\label{eq:delta_k_updated}
\end{equation} 

The shot noise is calculated as
\begin{equation}
\label{eq:SN_updated}
    P_{\rm SN}(k,\mu)=\dfrac{V}{M_t^2}\cdot\sum_i M_i^2\cdot e^{-\sigma_i^2\mu^2k^2}\ \ \ 
    ,\ \ \ \mu=k_{z}/k
\end{equation}
\begin{figure*}
    \centering
    \includegraphics[width=1.05\textwidth]{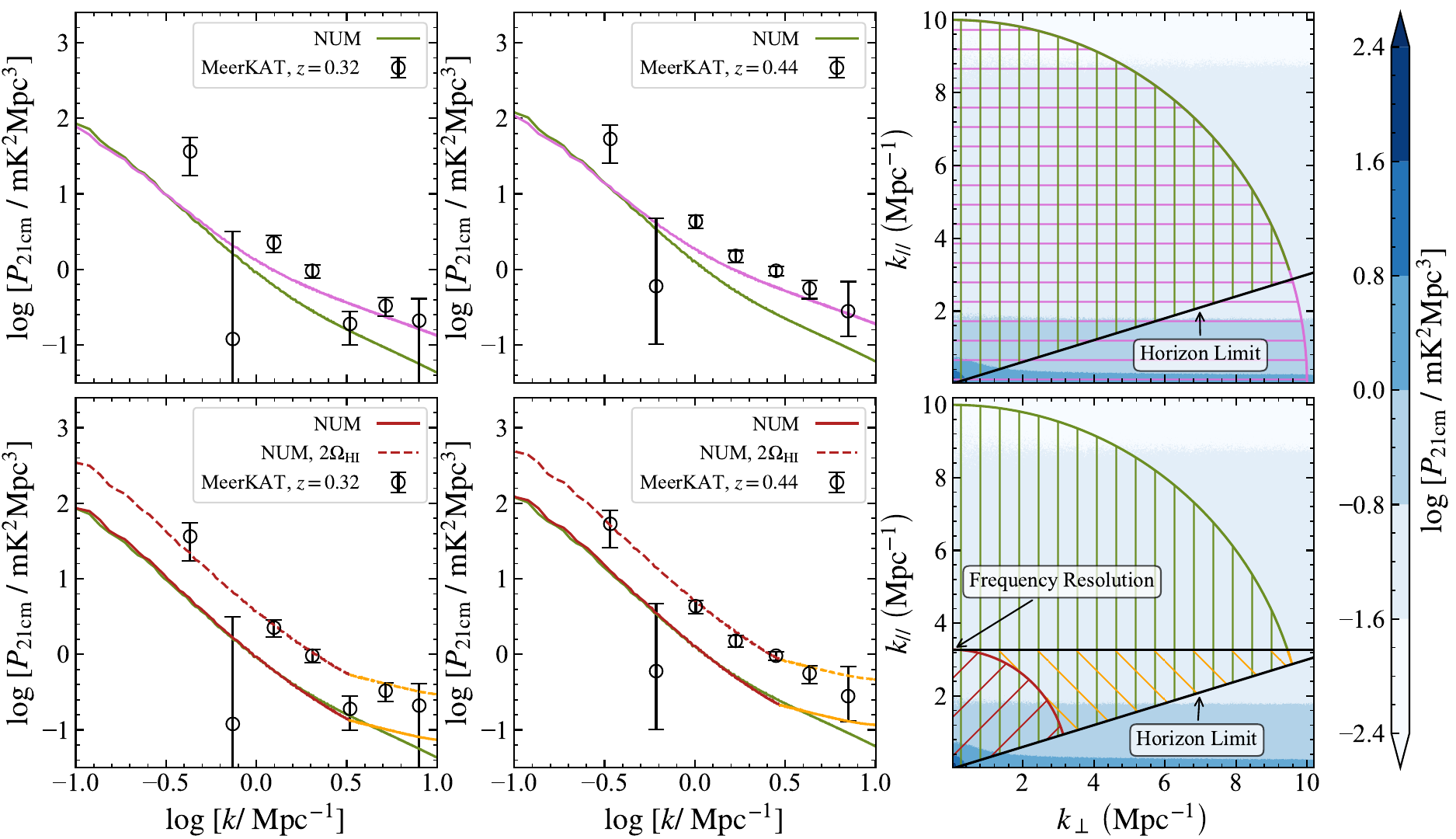}
	\caption{Top: How the horizon limit affects the measured 21cm power spectrum at $z=0.32,\ 0.44$. The black dots with error bars in the left and mid panels are the measurements, while the colored lines are the predicted 21cm power spectrum. Pink lines represent the original power spectrum, while green lines represent the power spectra that are truncated by the horizon limit. The black line labeled as ``Horizon Limit'' in the right panel is the one adopted by \citetalias{Paul2023}. Bottom: How the frequency resolution affects the measured 21cm power spectrum at $z=0.32,\ 0.44$. The black dots with error bars in the left and mid panels are the measurements, while the solid colored lines are the predicted 21cm power spectrum. The colored regions in the right panel present the cylindrical power spectrum that corresponds to the predicted 1D power spectra lines of the same color in left and mid panel. For comparison, predicted power spectra with larger $\Omega_{\rm HI}$ are also shown as the dashed colored lines.}
    \label{fig:ps_meerkat}
\end{figure*}
The theoretical power spectrum without shot noise term can be calculated using the simplest model as follows \citep{Peacock1992, Ballinger1996, Sarkar2019, Zhang2020}.
\begin{equation}
\begin{aligned}
    P^{(s)}(k,\mu)&=(1+\beta\mu^2)^2\cdot P_{\rm HI}(k)\cdot D_{\rm FoG}(k,\mu)
\end{aligned}
\label{eq: two-galaxy-term}
\end{equation}
where $\beta$ equals to $f/b_{\rm HI}$ with $f$ being the growth factor and $b_{\rm HI}$ being the bias of \hi; the FoG effect $D_{\rm FoG}$ is determined both by the peculiar velocity and the elongation due to the inner motions of sources. Since it is difficult to obtain the $D_{\rm FoG}$ theoretically, in this work, we calculated $P^{(s)}$ by subtracting the theoretical shot noise (red lines) from the simulated total power spectrum (black lines).

In particular, the power spectra without the shot noise and the shot noise term intersect at $k\approx 1\ \rm Mpc$, which coincides with the scale of the third measurement point of \hi auto power spectra by \citetalias{Paul2023}. As mentioned above, there are discrepancies between the \hi bias of different simulations and also measurements. Given the $k$ mode of the intersection, it is likely that different constant \hi biases would influence the power spectra at these scales. Therefore, changes in the \hi bias could impact the constraints on \hi abundance and parameters in the $W_{\rm 50}$ model. More tests using simulations with a higher \hi bias are necessary to further explore this issue in the future.

\subsubsection{Lower Limit of HIMF and the Shot Noise Term}

Taking into account the tight relation between $v_{\rm max}$ and $W_{\rm 50}$, we give an empirical formula for this relation to determine the profile width for \hi\ sources in our mock. However, in extragalactic \hi surveys such as HIPASS, ALFALFA and FASHI, it is more direct to obtain samples with combined information including \hi mass and corresponding $W_{\rm 50}$. Given the amplifying effect of squaring on large values, it is likely to reproduce the shot noise term even if sources with small \hi\ mass are not included.

We test how the lower limit of HIMF can impact the shot noise term using the theoretical method as Equation \ref{eq:SN_updated} and present the results in Figure \ref{fig:sn_frac}. Different colours represent different lower limits of the \hi mass. According to the middle panel, the knowledge of the most massive \hi sources ($M_{\rm HI} > 10^{9.2}\rm M_{\odot}$) can reproduce 99\% of the original shot noise.

In the middle panel of Figure \ref{fig:sn_frac}, the decreasing feature of the fractional line can be explained by the $W_{\rm 50}$-\hi mass relation depicted in the right panel. Massive \hi sources typically maintain large $W_{\rm 50}$, which causes suppression to occur at smaller $k$ modes, and vice versa. Consequently, the shot noise with larger lower limit appears to be less at large $k$ modes due to the absence of less massive \hi sources.

\section{Comparison with Observation}
\label{sec: obs}
In this section, we perform a comparative analysis between our model and the measurements reported by \citetalias{Paul2023} (see more details in Section \ref{sec: meerkat_data}). This analysis accounts for the observational settings including frequency resolution and the horizon limit, both of which significantly influence the observed 21cm power spectrum and help explain certain features shown in the measurement. Subsequently, simple fitting results are applied to provide initial interpretations of the measured data.

\subsection{Observational Settings}
\label{sec: observation settings}
\subsubsection{Horizon Limit}
\label{sec:horizon_eff}
Currently, the main challenge of \hi\ intensity mapping is extracting the signal from the foreground, which is several orders of magnitude brighter than the \hi\ signal. Fortunately, foregrounds such as Galactic synchrotron emission and radio galaxies exhibit smooth frequency characteristics. With a limited number of bright radio sources in the detection patches, \citetalias{Paul2023} adopted a foreground avoidance technique to separate the foreground from the signal, that is, by properly accounting for the horizon limit. According to \cite{Liu2014}, the horizon limit is taken into account using
\begin{equation}
    k_{\parallel}=\dfrac{xH_{0}E(z)\sin{\theta_0}}{c(1+z)}k_{\rm\perp}\,.
\label{eq:HL}
\end{equation}
where the $\theta_0$ is the angular extent of the MeerKAT dishes. By assuming an extreme case, \citetalias{Paul2023} adopted a horizon limit as $k_\parallel > 0.3k_{\rm\perp}$ to ensure the robustness of the measurements.

For comparison, this work also follows this limit to truncate the cylindrical 2D power spectrum to obtain the 1D power spectrum.
The upper panels of Figure \ref{fig:ps_meerkat} illustrate the impact of the horizon limit truncation on the measured 21cm power spectrum. The black dots with error bars represent the measurements reported by \citetalias{Paul2023}. Pink lines are the original prediction of the 21cm power spectrum with no observational settings considered, while green lines are the predicted 21cm power spectrum with a horizon limit considered, as shown by the diagram in the right panel. The horizon limit truncates the large parts of 2DPS and results in an obvious suppression on large $k$-modes and a subtle enhancement on small $k$-modes.

If we consider an extreme case without emission line profiles, the truncation by the horizon limit would not suppress the mock \hi power spectra on small scales. Thus, we suggest that the strength of this suppression depends not only on the size of the truncation, but also on how we model the $W_{\rm 50}$.

\subsubsection{Frequency Resolution}
\label{sec: FR}
Frequency resolution, given by the channel width of the receiver, is another observational effect that influences the measurements. For 21cm surveys, the frequency of the emission line, reveals the redshift of the source, so the frequency resolution represents the space resolution in the light-of-sight direction and therefore determines the largest $k_{\rm \parallel}$ that can be resolved as follows
\begin{eqnarray}
    k_{\parallel, \rm max}=\dfrac{\pi}{\Delta L}=\dfrac{100\pi h}{c}\cdot\dfrac{\nu_{\rm HI}}{\Delta\nu}\cdot\dfrac{E(z)}{(1+z)^2}
\end{eqnarray}
where the $\Delta L$ is the spatial resolution corresponding to the frequency resolution $209 \rm kHz$; $\Delta\nu$ is the frequency resolution, $\nu_{\rm HI}=1420\ \mathrm{MHz}$ is the frequency of 21cm signal; $h=H_0\ (km/s/{\rm Mpc})/100$ is the small Hubble constant; $E(z)=\sqrt{\Omega_m(1+z)^3+\Omega_{\Lambda}}$.

The bottom panels of Figure \ref{fig:ps_meerkat} illustrate the impact of frequency resolution on the measured 21cm power spectrum. The hard cut of frequency resolution in the cylindrical power spectrum shown in the right panel leads to a turning point in the corresponding 1D power spectrum in the left and mid panels. The red region in the right panel corresponds to the red part of predicted 1D power spectra in the left and mid panels, while the orange region corresponds to the orange 1DPS. The orange lines start from the third point from the bottom and help explain the suddenly slowing trend in the measurement. It can be inferred that if the frequency resolution gets higher, the turning points in the left and mid panel are pushed to the larger $k$ modes until the combined color lines converge to the green lines in Figure \ref{fig:ps_meerkat}.

The fiducial model, combined with observational settings, is depicted by the red plus orange solid lines. For comparison, the red plus orange dashed lines are also shown as the power spectra with double $\Omega_{\rm HI}$. The dashed lines closely match the measurements, indicating that while the amplitude of our models is much lower than that of the measurements, the trend of our prediction closely matches the measurements.

\subsection{Fittings Results}
\begin{figure}
        \centering
        \includegraphics[width=0.47\textwidth]{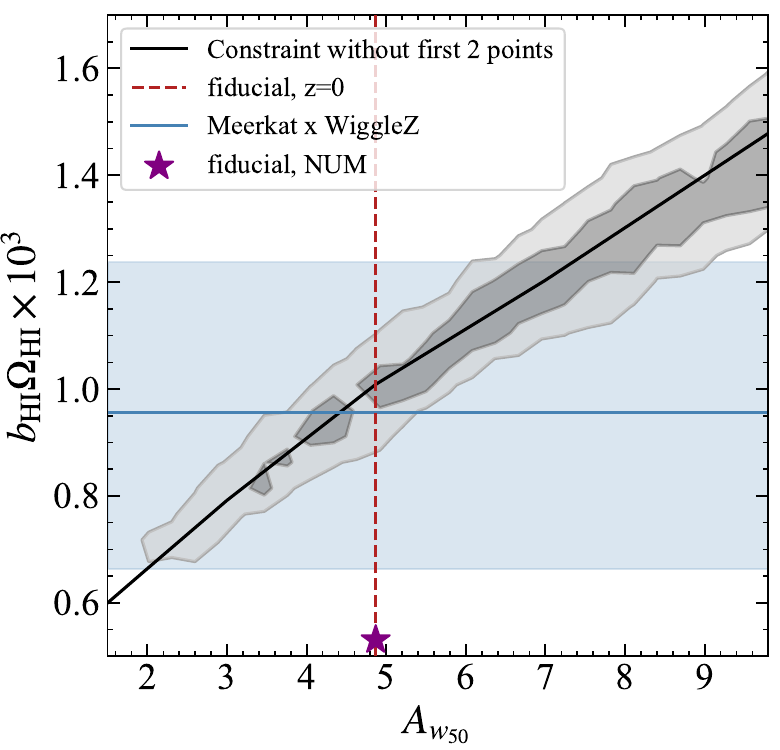}
        \caption{The parameter contour constrained by the MeerKAT auto-correlation measurement at $z\approx 0.44$. We illustrate the correlation between different parameters with 68\% (95\%) confidence level in grey (light-grey) contours. As a reference, the black solid line is obtained by fixing the parameter $A_{W_{\rm 50}}$ at different values and calculating the corresponding best $\Omega_{\rm HI}$. The firebrick dashed line represents our fiducial model, which is obtained from local universe measurements. The purple star is the $b_{\rm HI}\Omega_{\rm HI}\times 10^3$ embedded in the NUM model. The blue line and shadow area denote the estimation and error bar, respectively, derived from MeerKAT x WiggleZ cross correlation measurements at $z\approx 0.43$.}
        \label{fig: contour}
    \end{figure}

Given the absence of $W_{\rm 50}$ measurements beyond $z=0.2$ and the consequent uncertainty in the $W_{\rm 50}$ model, we treat both $A_{W_{\rm 50}}$ and $\Omega_{\rm HI}$ as free parameters in our mock. To further understand and translate the measurements into constrained parameters within the model, we constrain the two parameters using the measurement reported by \citetalias{Paul2023} at $z=0.44$ via the Monte Carlo Markov Chain (MCMC) sampling method. Assuming a $\chi$-square likelihood, we fit the model parameters using the EMCEE package \citep{Foreman2013}. We employ eight random walkers for the two parameters and set 3000 steps for each walker to ensure convergence. We drop the first 91 steps. Gaussian prior is established for $\Omega_{\rm HI}$ according to the measurements made by \cite{Rao2017} at $z\approx0.46$, and the flat prior ranging from 0 to 10 is established for $A_{W_{\rm 50}}$. The grey (light-gray) contour in Figure \ref{fig: contour} illustrates a strong degeneracy between $A_{W_{\rm 50}}$ and $\Omega_{\rm HI}$. The likelihood of samples in the grey contour is very close, therefore the best-fit parameter is not shown here.

To ensure robustness, we exclude the first two measurement data points with large error bars reported by \citetalias{Paul2023} from the MCMC analysis. According to our test (not shown here), the contour shape is barely affected even with the first two points being included. For comparison, the solid black line represents the $\Omega_{\rm HI}$ values obtained by fixing $A_{W_{\rm 50}}$ at different values and computing the corresponding best-fitting $\Omega_{\rm HI}$. Additionally, the red dashed line represents our fiducial model, obtained from fittings to the $W_{\rm 50}$ measurements in the local universe(see Section \ref{sec:w50}).

The constant \hi\ bias in our simulation is fixed to $b_{\rm HI} = 0.82$ at $z\approx0.44$, and we include it into the $\Omega_{\rm HI}b_{\rm HI}$ parameter, shown as the label of the y-axis of the figure. Given the existing disagreement regarding \hi\ bias and the possible impacts of $b_{\rm HI}$ on the mock $P_{\rm 21cm}$, the $\Omega_{\rm HI}b_{\rm HI}$ is presented to mitigate the potential discrepancies related to \hi\ bias. 

The constraint from fitting the measurements of \citetalias{Paul2023} is shown as the grey contours. It is clear that $\Omega_{\rm HI}b_{\rm HI}$ is strongly coupled with $A_{W_{\rm 50}}$. Without an accurate estimate of $A_{W_{\rm 50}}$, it is difficult to place tight constraints on $\Omega_{\rm HI}b_{\rm HI}$. The constraints of $\Omega_{\rm HI}b_{\rm HI}$ seem to be larger than the fiducial value of the NUM model (shown as the purple star) by a factor of 2 at the fiducial $A_{W_{\rm 50}}$. It could be possibly caused by the uncertainties of the observational measurements and the underestimate of $\Omega_{\rm HI}b_{\rm HI}$ in the NUM model by fitting the $\Omega_{\rm HI}$ measurements in literature. More measurements are needed to further confirm this inconsistency and explore the underlying causes.

For comparison, we also include another observational measurements in the figure, i.e., the MeerKAT radio telescope x WiggleZ Dark Energy Survey (abbreviated as ``MeerKAT x WiggleZ'' in the figure) cross-correlation measurements \citep{Cunnington2023} at an effective scale of $k_{\rm eff} \approx 0.13 h\rm Mpc^{-1}$. The cross-correlation applied the \hi data measured by the single-dish modes of MeerKAT, so it provides no constraint for the $A_{W_{\rm 50}}$ parameter due to the large detected scale. Our constraint result aligns closely with the cross-correlation measurements, as evidenced by the proximity of our fiducial model to the intersection of the blue and black lines. This is because, in any case, we do not expect $W_{\rm 50}$ to vary dramatically when the redshift changes only from 0.0 to 0.44.

With an increasing number of measurements of the \hi\ power spectrum on non-linear scales, $P_{\rm 21 cm}$ shows promise as a statistic for imposing additional constraints on both $\Omega_{\rm HI}$ and $W_{\rm 50}$ or other profile width quantities like $W_{\rm 20},W_{\rm 10}$ in future studies.

\section{Conclusion}
\label{sec: conclusion}

In this paper, we present a simulated-based framework to predict the measured 21cm power spectrum on non-linear scales and explore impacts of different factors on the power spectrum.

We validate the capability of the measured data to resolve galactic and subhalo structures both perpendicular to and along the line of sight (LOS), as depicted in Figure \ref{fig:which_one}. We outline the detailed methodology employed in this study, focusing on generating the \hi\ density field from catalog data and deriving corresponding 1D power spectra. We then introduce a novel approach for simulating the emission line profile for each \hi\ source. Our analysis reveals that variations in the shape of the emission line profile have subtle effects on the \hi\ power spectra, as illustrated in Figure \ref{fig:vf_function}, while the profile width, represented by the Full Width at Half Maximum ($W_{\rm 50}$), emerges as a critical factor.

To determine $W_{\rm 50}$ for each source, we propose an empirical model that fits well the width function of $W_{\rm 50}$ measured by the ALFALFA survey. This model expresses $W_{\rm 50}$ as a double power law function of the maximum velocity ($v_{\rm max}$) of subhalos. This choice stems from the expectation that the widths of gas velocity profiles should closely track the velocity profiles of their host subhalos. The model involves four free parameters: $A_{W_{\rm 50}}$, $V_{0}$, $\alpha$, and $\beta$. Among these parameters, the \hi\ power spectrum is particularly sensitive to variations in $A_{W_{\rm 50}}$. Throughout our analysis, we adopt the best-fit values of these parameters as the fiducial model.

To provide an interpretation of the measurements conducted by \citetalias{Paul2023}, we consider the observational settings employed in their study, including the horizon limit and frequency resolution. In Figure \ref{fig:ps_meerkat}, we visualize how these settings impact the \hi\ power spectrum. The horizon limit notably suppresses the power spectrum on small scales, while the frequency resolution introduces a turning point at the corresponding $k$ mode. Due to the considerable uncertainty of $\Omega_{\rm HI}$ and $W_{\rm 50}$ at $z=0.44$, we treat $A_{W_{\rm 50}}$ and $\Omega_{\rm HI}$ as free parameters and constrain them using the measurements. Shown as Figure \ref{fig: contour}, the constraint result reveals a strong degeneracy between these parameters. The line slope between $A_{W_{\rm 50}}$ and $\Omega_{\rm HI}$ remains unchanged, regardless of whether the first two measurement points are included. Remarkably, our constrained results align with measurements of $b_{\rm HI}\Omega_{\rm HI}$ obtained through cross-correlation analyses.

\section*{Acknowledgements}

We would like to thank Zhaoting Chen, Wenxiu Yang, Aishrila Mazumder, Sourabh Paul and Keith Grainge for helpful discussion; Yu Lei, Meng Yang for useful discussion regarding the \hi\ emission line; Zipeng Liu, Jiajun Zhang and Zhenyuan Wang for useful discussions regarding the signal processing; Tao Jing and Weicheng Zang for useful discussion regarding the MCMC sampling method; Marta Spinelli for useful discussions concerning the \hi\ bias measurement; Jonghwan Rhee for kindly providing the \hi\ abundance measurement data.

ZL and YM are supported by the National SKA Program of China (grant No.~2020SKA0110401) and NSFC (grant No.~11821303). LW is a UK Research and Innovation Future Leaders Fellow [grant MR/V026437/1]. HG is supported by the National SKA Program of China (grant No. 2020SKA0110100), the CAS Project for Young Scientists in Basic Research (No. YSBR-092) and the science research grants from the China Manned Space Project with NO. CMS-CSST-2021-A02. SC is supported by a UK Research and Innovation Future Leaders Fellowship
grant [MR/V026437/1]. 
We acknowledge the Tsinghua Astrophysics High-Performance Computing platform at Tsinghua University and the Core Facility for Advanced Research Computing at the Shanghai Astronomical Observatory for providing computational and data storage resources that have contributed to the research results reported within this paper.

\section*{Data Availability}

The catalogs used in this study will be shared on reasonable request with the corresponding authors. Additional derived data are available in the article.



\bibliographystyle{mnras}
\bibliography{ref} 




\bsp	
\label{lastpage}
\end{document}